\documentclass[twocolumn,showpacs,preprintnumbers,amsmath,amssymb,prl,epsf,superscriptaddress]{revtex4}

\usepackage{graphicx}
\begin{document}
\title{Chirped Biphotons and Their Compression in Optical Fibres}
\author{G.~Brida}
\affiliation{Istituto Nazionale di Ricerca Metrologica, Strada delle
Cacce 91, 10135 Torino, Italy}
\author{M.~V.~Chekhova}
\affiliation{Department of Physics, M.V.Lomonosov Moscow State
University,\\  Leninskie Gory, 119992 Moscow, Russia}
\affiliation{Istituto Nazionale di Ricerca Metrologica, Strada delle
Cacce 91, 10135 Torino, Italy}
\author{I.~P.~Degiovanni}
\affiliation{Istituto Nazionale di Ricerca Metrologica, Strada delle
Cacce 91, 10135 Torino, Italy}
\author{M.~Genovese}
\affiliation{Istituto Nazionale di Ricerca Metrologica, Strada delle
Cacce 91, 10135 Torino, Italy}
\author{G.~Kh.~Kitaeva}
\affiliation{Department of Physics, M.V.Lomonosov Moscow State
University,\\  Leninskie Gory, 119992 Moscow, Russia}
\author{A.~Meda}
\affiliation{Istituto Nazionale di Ricerca Metrologica, Strada delle
Cacce 91, 10135 Torino, Italy}
\author{O.~A.~Shumilkina}
\affiliation{Department of Physics, M.V.Lomonosov Moscow State
University,\\  Leninskie Gory, 119992 Moscow, Russia}

\begin{abstract}
\begin{center}\parbox{14.5cm}
{We show that broadband biphoton wavepackets produced via
Spontaneous Parametric Down-Conversion (SPDC) in crystals with
linearly aperiodic poling can be easily compressed in time using the
effect of group-velocity dispersion in optical fibres. This result
could foster important developments in quantum metrology and
lithography.}
\end{center}
\end{abstract}
 \maketitle \narrowtext
\vspace{-10mm}

One of the central problems in quantum optics is generation of
nonclassical light with given spectral and spatiotemporal
properties. In particular, for the needs of quantum metrology and
quantum lithography it is important to obtain two-photon wavepackets
with small correlation times. Such wavepackets should naturally
manifest a broad frequency  spectrum. Several ideas have been put
forward in this direction, all based on two-photon states produced
via Spontaneous Parametric Down-Conversion (SPDC). Among them, one
can mention prisms or diffraction gratings introducing a frequency
chirp~\cite{Valencia}, SPDC  in aperiodically poled
crystals~\cite{Carrasco,Harris, Nasr}, and SPDC in crystals with
temperature gradients~\cite{Kalashnikov}. However, a broad spectrum
of two-photon light does not necessarily imply small correlation
times, although the inverse is
true~\cite{spectron,spectronPRL,Strekalov}. This is similar to the
fact that a broadband pulse does not have to be short in time,
although the spectrum of a short pulse is always broad. The spectrum
broadening introduced in Refs~\cite{Valencia,Kalashnikov,Nasr} is in
fact inhomogeneous; as a result, the two-photon spectral amplitude
in all these cases has a phase depending nonlinearly on the
frequency. This phase (a frequency chirp \cite{chirp}) makes
two-photon wavepackets not Fourier transform-limited. Therefore,
they are not short in time despite their broad frequency spectrum.
As it was mentioned in Ref.~\cite{Harris}, time compression of such
two-photon wavepackets requires compensation for their frequency
chirp. At the same time, the way to eliminate the chirp was not
specified.

In this paper we show that, under certain conditions, biphoton
wavepackets can be made nearly Fourier transform-limited and hence
compressed by injecting one of the photons of a pair in a standard
optical fibre and exploiting the effect of group-velocity dispersion
(GVD). No specially engineered fibres (for instance, with negative
GVD) are necessary. This suggests an easy way of achieving extremely
short correlation times for two-photon light.

Consider  generation of two-photon light via spontaneous parametric
down-conversion (SPDC) from a cw pump in an aperiodically poled
crystal.  From the viewpoint of applications and for simplifying the
calculation, it is convenient to assume that signal and idler
photons are distinguishable, due to either frequency nondegenerate
or type-II phase matching. Below, we consider phase matching to be
type-II, collinear, and frequency degenerate, with idler
(extraordinary) and signal (ordinary) radiations centered at
frequency $\omega_{0}$. The two-photon state can be written as
\begin{equation}
|\psi\rangle=\int\hbox{d}\Omega F(\Omega)|\omega_{0}-\Omega
\rangle_{i} |\omega_{0}+\Omega\rangle_{s}, \label{state}
\end{equation}
where $|\omega\rangle_{i (s)}$ denotes the idler (signal) photon
state with frequency $\omega$. The two-photon spectral amplitude
(TPSA) $F(\Omega)$ determines all spectral and temporal properties
of two-photon light. In particular, its squared module gives the
frequency spectra of signal and idler radiation,
$I_{s,i}(\omega)\propto|F(\omega-\omega_{0})|^2$. Its Fourier
transform can be called time two-photon amplitude (TTPA)
\cite{Klyshko},
\begin{equation}
F(\tau)=\int d\Omega e^{i\Omega\tau}F(\Omega),
\label{Fourier}
\end{equation}
its squared module giving the second-order Glauber's correlation
function~\cite{spectron,spectronPRL}: $G^{(2)}(\tau)=|F(\tau)|^2$.
The TPSA is determined by the distribution of the quadratic
nonlinearity $\chi(z)$ along the
crystal~\cite{Klyshko,Harris,Kitaeva}:
\begin{equation}
F(\Omega) \propto \int_{-L}^{0} \mathrm{d}z ~\chi(z)
e^{i(k_i+k_s-k_p)z}. \label{TPSA}
\end{equation}
Here, $L$ is the crystal length and $k_i, k_s, k_p$ are wavevectors
of the idler, signal, and pump waves, respectively. Let the spatial
dependence of the quadratic nonlinearity be
$\chi(z)=\chi_0e^{iK(z)(z+L/2)}$, where the inverse grating vector
$K$ has a linear dependence on the coordinate, $K(z)=K_0-\alpha
(z+L/2)$ \cite{Nasr,Harris,chirp}, and $K_0$ provides
quasiphasematching: $k_i(\omega_{0})+k_s(\omega_{0})-k_p+K_0=0$ at
the center of the crystal. It is convenient to expand the
wavevectors around the exact quasiphasematching frequency:
\begin{eqnarray}
\nonumber
k_i=k_i(\omega_{0})-k'_i\Omega+\frac{1}{2}k''_i\Omega^2,\\
k_s=k_s(\omega_{0})+k'_s\Omega+\frac{1}{2}k''_s\Omega^2.
\label{expansion}
\end{eqnarray}
Here, $k'_{i,s}$ and $k''_{i,s}$ are the first and second
derivatives of the dispersion law evaluated at $\omega_0$, related
to the group velocity and group velocity dispersion (GVD),
respectively.

Denoting $D\equiv k'_s-k'_i$ and
$\kappa\equiv\frac{1}{2}(k''_i+k''_s)$, we obtain the TPSA in the
form
\begin{equation}
F(\Omega)\propto
e^{-iD\Omega\frac{L}{2}-i\kappa\Omega^2\frac{L}{2}}\int_{-\frac{L}{2}}^{\frac{L}{2}}\hbox{d}\xi~
\chi_0 e^{i(D\Omega+\kappa\Omega^2)\xi-i\alpha \xi^2}, \label{TPSA1}
\end{equation}
where $\xi=z+L/2$.

Suppose that the spectrum is not too broad compared to the
difference of group velocities of the signal and idler radiation, so
that the condition
\begin{equation}
\left|\frac{\kappa\Omega}{D} \right| \ll 1 \label{cond1}
\end{equation}
holds true. Then the wavevector mismatch can be written  up to
linear terms in frequency detuning $\Omega$, and the TPSA becomes
\begin{equation} F(\Omega)\propto
e^{-iD\Omega\frac{L}{2}}\int_{-\frac{L}{2}}^{\frac{L}{2}}\hbox{d}\xi
~\chi_0 e^{iD\Omega \xi-i\alpha \xi^2}, \label{TPSA2}
\end{equation}
which yields, similarly to Ref.~\cite{Harris},
\begin{eqnarray}
\nonumber
F(\Omega)\propto \hbox{exp}\{-iD\Omega\frac{L}{2}+i\frac{D^2\Omega^2}{4\alpha}\}\\
\times\left\{\hbox{erf}\left[\sqrt{\frac{i}{\alpha}}\frac{L\alpha-D\Omega}{2}\right]
+\hbox{erf}\left[\sqrt{\frac{i}{\alpha}}\frac{L\alpha+D\Omega}{2}\right]\right\}.
\label{TPSA3}
\end{eqnarray}

In the case of large aperiodicity $\alpha$, the spectral amplitude
has a rectangular shape. Indeed, let us introduce the `rectangle
function' $\Pi(x,a,b)\equiv 1$ for $a\le x\le b$ and
$\Pi(x,a,b)\equiv 0$ otherwise. Then, rewriting the integral in
(\ref{TPSA2}) in terms of the rectangle function and applying the
convolution theorem, we get
\begin{equation}
F(\Omega)\propto
e^{-iD\Omega\frac{L}{2}+i\frac{D^2\Omega^2}{4\alpha}}\int_{-\infty}^{\infty}\hbox{d}x~\hbox{sinc}\{Dx\frac{L}{2}\}
e^{i\frac{D^2(x^2-2\Omega x)}{4\alpha}}, \label{TPSA4}
\end{equation}
where $\hbox{sinc}(x)\equiv \sin(x)/x$. The first exponential term
in the integral, $\hbox{exp}\{i\frac{D^2x^2}{4\alpha}\}$, can be
omitted if the typical scale of its variation is much larger than
the sinc-function width, $\pi/DL$. This is the case if the
aperiodicity is large enough,
\begin{equation}
|\alpha|\gg \frac{\pi^2}{4L^2}. \label{large_chirp}
\end{equation}
Note that condition (\ref{large_chirp}) is well satisfied in
Refs.~\cite{Harris} and ~\cite{Nasr}. Then, the integral in
Eq.~(\ref{TPSA4}) becomes
\begin{equation}
F(\Omega)\propto
e^{-iD\Omega\frac{L}{2}+i\frac{D^2\Omega^2}{4\alpha}}\Pi(\Omega,-\frac{\alpha
L}{D},\frac{\alpha L}{D}). \label{TPSArect}
\end{equation}

We see that the spectrum of SPDC in a crystal with linear $K(z)$
dependence and large $\alpha$ is a rectangular function of width
$\Delta\Omega=\frac{2\alpha L}{D}$. The condition
(\ref{large_chirp}) means physically that the aperiodicity should
induce a substantial spectrum broadening. Now we can explicitly
write the condition (\ref{cond1}) for the GVD of the nonlinear
crystal to be negligible:
\begin{equation}
\left|\frac{\kappa L\alpha}{D^2}\right|\ll 1. \label{cond2}
\end{equation}
For given $\alpha$ and $L$, this condition is realized if the
separation of signal (ordinary) and idler extraordinary group
velocities is large enough.

Increasing the aperiodicity $\alpha$, one can make the spectrum as
broad as desired. At the same time, this does not make the TTPA
(\ref{Fourier}) short in time because, due to the nonlinear
frequency-dependent phase factor in (\ref{TPSA3}), the TPSA is not
Fourier transform-limited. Here we would like to stress that because
the squared module of TTPA is the second-order Glauber's correlation
function, its width gives the correlation time of the biphoton, i.e.
the biphoton entanglement time~\cite{spectron,spectronPRL}. It is
this time that is important for two-photon effects such as
two-photon absorption, two-photon ionization, or up-conversion, and
which can be measured for two-photon light using these
techniques~\cite{Silberberg}. At the same time, coherence time of
biphoton light is defined as the width of the first-order Glauber's
correlation function, which is the Fourier transform of the spectrum
$|F(\Omega)|^2$. This is why coherence time does not depend on the
phase factor in (\ref{TPSA3}) and it is given by the inverse width
of the spectrum~\cite{Nasr}.

The Fourier transform of TPSA (\ref{TPSA2}) is easily obtained by
introducing the rectangle function under the integral, extending the
integration to infinite limits and using the convolution theorem. As
a result, we get
\begin{equation}
F(\tau)\propto e^{-i\alpha(\frac{L}{2}-\frac{\tau}{D})^2}\Pi(\tau,0,DL),
\label{TTPA}
\end{equation}
with the amplitude being the same as in the case of a bulk or
periodically poled crystal of length $L$. This means that the TTPA
of an aperiodiocally poled crystal is as broad as in the absence of
the aperiodicity $\alpha$.

Consider now propagation of the extraordinary photon of the biphoton
field through an optical fibre of length $l$ with the inverse group
velocity given by $k_f'\equiv\frac{dk}{d\omega} |_{\omega=\omega_0}$
and the GVD given by $\kappa_f \equiv \frac{1}{2}
\frac{d^2k}{d\omega^2} |_{\omega=\omega_0}$. Propagation through
such a fibre leads to a phase factor $\hbox{exp}\{i(k_f' \Omega
+\kappa_f \Omega^2)l\}$ in the two-photon spectral
amplitude~\cite{spectron,spectronPRL}. The first term in the phase
is linear in frequency and hence only shifts the two-photon
wavepacket in time. The second term, being quadratic in frequency,
can compensate for the TPSA chirp. This will happen under the
condition
\begin{equation}
\kappa_f l+\frac{D^2}{4\alpha}=0. \label{compens}
\end{equation}
For a fibre with positive GVD, this condition can be satisfied for
negative $\alpha$, i.e., for the case where the poling period
reduces along the pump propagation through the crystal.

In the case of large aperiodicity $\alpha$ satisfying
(\ref{large_chirp}), the resulting TTPA can be calculated
analytically as the Fourier-transform of expression (\ref{TPSArect})
with the quadratic phase term removed. Clearly, it has the form of a
sinc-function with the width being almost $\alpha L^2$ times
narrower than a periodically poled sample of the same length.

Figure 1 shows the spectrum of SPDC radiation calculated for the
case of aperiodically poled KTP crystal with $L=0.8$ cm,
$K_0=2441.8$ cm$^{-1}$ and $\alpha=1200$ cm$^{-2}$, which
corresponds (using the first-order quasi-phasematching) to the
poling period varying from $18.47$ to $42.40$ $\mu$m. The pump at
$458$ nm is y-polarized, as well as the idler radiation, and all
three wavevectors, as well as the inverse grating vector, are
directed along x. The dispersion dependencies are given by Sellmeier
equations from Ref.~\cite{Sellmeier}, without any additional
assumptions.
\begin{figure}
\includegraphics[width=0.3\textwidth]{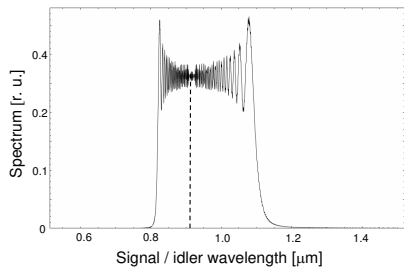}
\caption{(color online) Calculated spectrum of the signal/idler
radiation (solid line). Dashed line: the corresponding spectrum for
a crystal with $K=K_0$, reduced in the vertical scale by $2~10^2$.}
\end{figure}
One can see that the spectrum of the biphoton field is quite broad
(from 800 to 1200 nm) and has nearly rectangular shape. This is due
to the fact that condition (\ref{large_chirp}) is fulfilled very
well. In fact, the aperiodicity leads to the spectral broadening of
more than two orders of magnitude. For comparison, the same figure
shows the spectra of signal and idler radiation for a crystal with
the same length but constant poling period $K=K_0$, corresponding to
$\alpha=0$. Condition (\ref{cond2}) is reasonably satisfied, since
$|\frac{\kappa L\alpha}{D^2}|\approx 0.16$.

Figure 2 shows the second-order correlation function calculated as
the squared module of the Fourier transform of expression
(\ref{TPSA3}). For comparison, second-order correlation function of
a crystal with constant poling period is plotted in the same graph.
Clearly, both distributions have the same width, which means that
the biphoton with the broadened spectrum has the same correlation
time as the narrowband biphoton, i.e., it is not Fourier
transform-limited.
\begin{figure}
\includegraphics[width=0.3\textwidth]{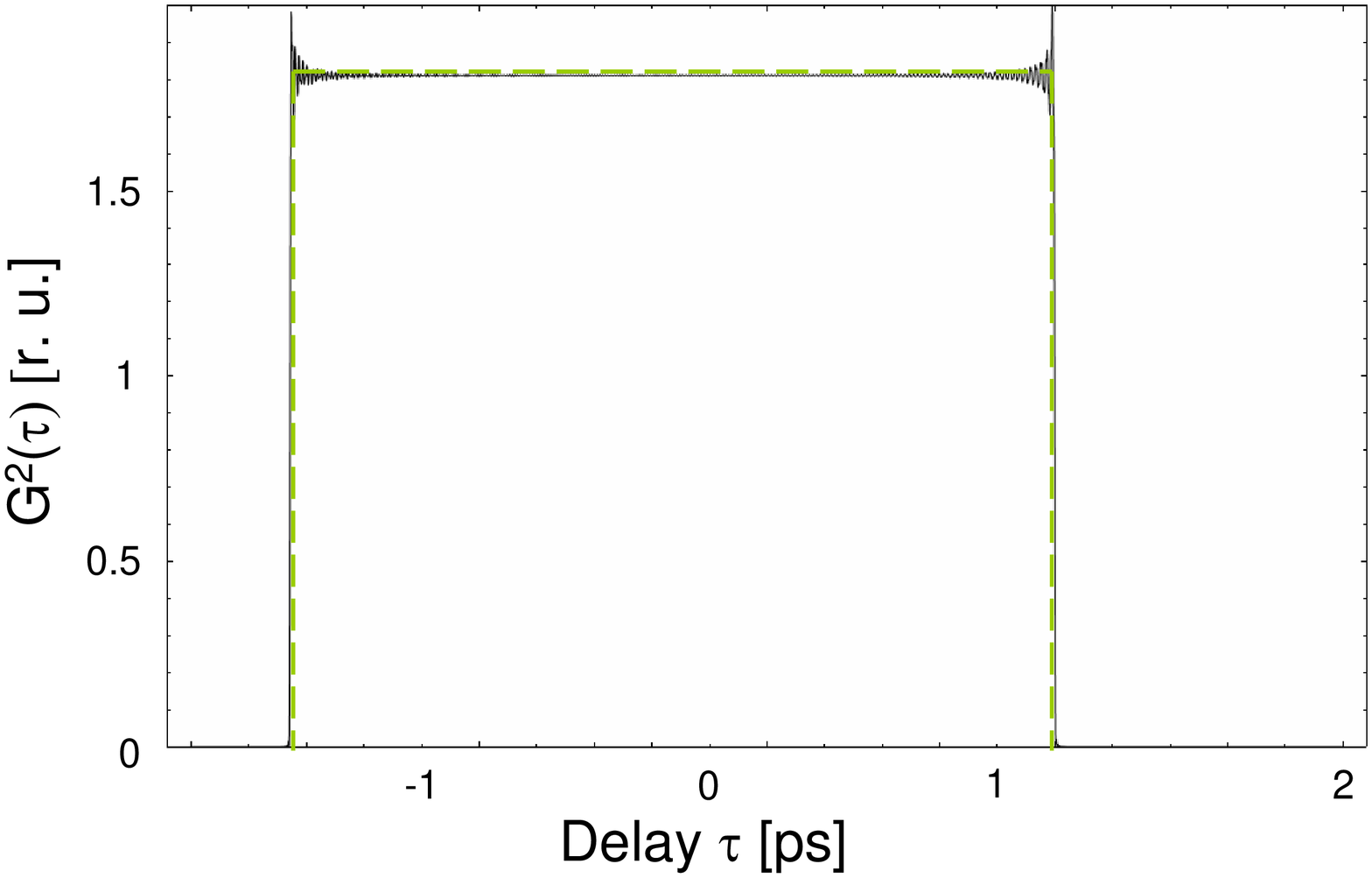}
\caption{(color online) Calculated second-order correlation function
of the biphoton in the case of aperiodic poling (black solid line)
and periodic poling (green dashed line).}
\end{figure}
Evolution of the correlation time in an optical fibre is
demonstrated in Fig.3 for the cases of $\alpha>0$ and $\alpha<0$,
when only the idler photon is transmitted through the fibre. We see
that at $\alpha>0$, propagation through the fibre only broadens the
biphoton wavepacket, while in the case $\alpha<0$, which is achieved
by simply exchanging the input and output faces of the crystal, the
biphoton is compressed. For the calculation, we used the GVD of bulk
fused silica~\cite{bulk}, because the waveguide contribution into
GVD far from zero dispersion point is negligibly
small~\cite{Agrawal}. The value of $\kappa_f$ we used in the
calculation was $1.359\cdot 10^{-28}\frac{\hbox{s}^2}{\hbox{cm}}$.
\begin{figure}
\includegraphics[width=0.3\textwidth]{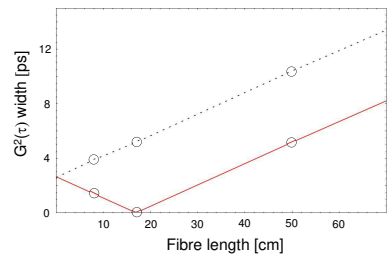}
\caption{(color online) TTPA width at the output of the fibre versus
the fibre length for $\alpha>0$ (red line) and $\alpha<0$ (grey
line). Circles denote the points for which the shapes of the
second-order correlation function are shown in Fig.4.}
\end{figure}
The largest compression of the biphoton wavepacket is achieved at
the fibre length $l=16.927$ cm. The second-order correlation
function in this case has a typical shape of squared sinc-function
with the FWHM equal to $12$ fs. At other lengths of the fibre, the
two-photon wavepacket is broader. As the length of the fibre
increases, the shape of $G^{(2)}(\tau)$ becomes similar to the shape
of the spectrum, an effect that was studied in detail in
Refs.~\cite{spectron,spectronPRL}. Fig.4 shows the shapes of
$G^{(2)}(\tau)$ after the biphoton propagation through fibres of
different length; the corresponding points are shown in Fig.3.

\begin{figure}
\includegraphics[width=0.5\textwidth]{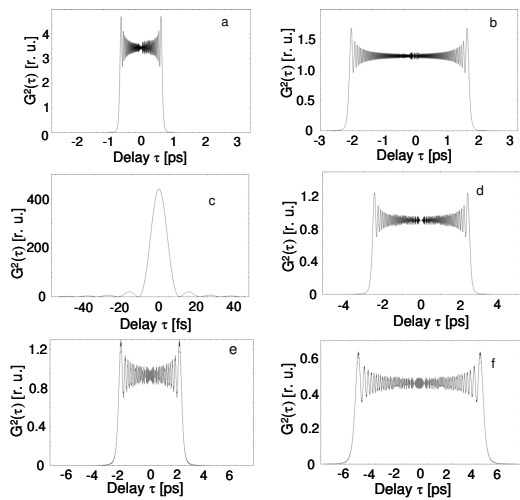}
\caption{(color online) Second-order Glauber's correlation function
of the biphoton with $\alpha<0$ (a,c,e) and $\alpha>0$ (b,d,f),
after its propagation through a fibre of length 8 cm (a,b), 16.927
cm (c,d), 50 cm (e,f)}.
\end{figure}

The model we have been using so far is based on the linear
dependence of the wavevector mismatch on the frequency detuning from
exact phasematching. This is valid for type-II or frequency
non-degenerate type-I SPDC \cite{footnote}, under the condition
(\ref{cond2}). Similarly, the dispersion law of the fibre in our
consideration was described by a quadratic dependence, i.e.,
third-order GVD of the fibre was neglected. In order to see the
effect of higher-order GVD terms, we have performed exact numerical
calculation for the same case as considered above. The results
(Fig.5) show that the effect of compression is slightly reduced but
there still remains a significant narrowing of the TTPA, useful for
applications. Even without any optimization, the correlation time is
reduced by more than an order of magnitude. An exhaustive study of
this effect and a search for optimized parameters of both the
crystal and the fibre will be presented in the nearest
future~\cite{meda}.
\begin{figure}
\includegraphics[width=0.3\textwidth]{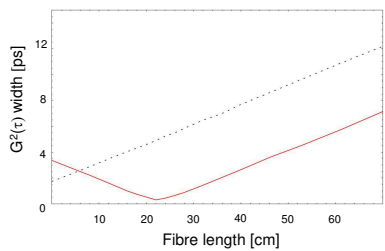}
\caption{(color online) TTPA width at the output of the fibre versus
the fibre length for $\alpha>0$ (red line) and $\alpha<0$ (grey
line) calculated without neglecting higher-order GVD terms.}
\end{figure}

An interesting feature of Fig.5 is that at the output of the
crystal, TTPA widths for the cases $\alpha>0$ and $\alpha<0$ are
different. This can be explained as follows. Although at the center
of the crystal signal and idler photons are generated with the same
frequencies, biphotons generated at the back face are
non-degenerate. Due to GVD, there is a delay accumulated between the
photons of a pair in the course of its propagation through the
crystal. At $\alpha<0$, this delay is compensated by the one
appearing due to birefringence and at $\alpha>0$, both delays add
up.

In conclusion, a biphoton whose spectrum is broadened due to a
linear aperiodicity of the crystal poling can be compressed in time
using normal GVD of an optical fibre. To describe the compression,
it is sufficient to take into account first-order terms in the
crystal dispersion dependence and second-order terms in the fibre
dispersion dependence. Exact calculation shows that higher-order
terms reduce the compression but the effect is still present. It is
worth mentioning an interesting result: the two-photon correlation
time as well as its evolution due to the propagation of the biphoton
through an optical fibre strongly depend on the position of the
crystal.

This work has been supported in part by the joint grant
RFBR-Piedmont 07-02-91581-ASP, RFBR grant 09-02-92003-NNS, MIUR
(PRIN 2007FYETBY), Regione Piemonte (E14),  "San Paolo foundation",
NATO (CBP.NR.NRCL 983251), RFBR 08-02-00555a and the Russian Program
for Scientific Schools Support, grant \# NSh-796.2008.2. M.~V.~Ch.
also acknowledges the support of the Lagrange project of the CRT
Foundation for her stay at INRIM.


\begin{references}
\bibitem{chirp} Sometimes this dependence is called a linear chirp
of the inverse grating vector; however, below we keep the term
'chirp' for the frequency dependence of the phase.
\bibitem{footnote} Note that the same technique will be valid if the spectral
broadening of two-photon light is caused by any physical mechanism
leading to a linear variation of the wavevector mismatch along the
crystal, in particular, by temperature gradients.
\end{references}
\end{document}